\documentclass[aps,pra,floatfix,showpacs,twocolumn,tightenlines,superscriptaddress]{revtex4}
\usepackage{amsmath}
\usepackage{graphicx}
\usepackage[hypertex]{hyperref}
\usepackage[usenames]{color}
\usepackage{bm}
\usepackage{amsfonts}
\usepackage{amsthm}
\usepackage{amssymb}
\usepackage{subfigure}
\usepackage{eucal}
\usepackage{dsfont}
\newtheorem{theorem}{Theorem}[section]

\newtheorem{lemma}{Lemma}[section]
\newtheorem{proposition}{Proposition}[section]
\newtheorem{conjecture}{Conjecture}[section]

\def\bra#1{\mathinner{\langle{#1}|}}
\def\ket#1{\mathinner{|{#1}\rangle}}
\def\defeq{\mathrel{\mathop:}=}
\def\ro{\mbox{\boldmath $\rho$}}
\def\sig{\mbox{\boldmath $\sigma$}}
\def\btau{\mbox{\boldmath $\tau$}}
\def\bLamb{\mbox{\boldmath $\Lambda$}}
\def\rpar{\right)}
\def\lpar{\left(}
\def\rbk{\right]}
\def\lbk{\left[}
\def\rbr{\right\}}
\def\lbr{\left\{}
\def\lb{\label}
\def\indn{\mbox{\tiny $N$}}
\def\idhs{\mbox{\tiny ${\rm HS}$}}
\def\iqcb{\mbox{\tiny ${\rm QCB}$}}
\newcommand{\be}{\begin{equation}}
\newcommand{\ee}{\end{equation}}
\newcommand{\brr}{\begin{eqnarray}}
\newcommand{\err}{\end{eqnarray}}
\newcommand{\nn}{\nonumber}
\newcommand{\bd}{\begin{displaymath}}
\newcommand{\ed}{\end{displaymath}}

\newcommand{\bib}{\bibitem}
\newcommand{\bfig}{\begin{figure}}
\newcommand{\efig}{\end{figure}}
\newcommand{\ie}{i.e.}
\newcommand{\eg}{e.g.}
\newcommand{\D}{\mathcal{D}}
\newcommand{\E}{\mathcal{E}}
\newcommand{\F}{\mathcal{F}}
\newcommand{\Fn}{\mathcal{F}_{\indn}}
\DeclareMathOperator{\tr}{Tr}
\begin{document}
\title{Alternative fidelity measure between quantum states}
\author{Paulo E. M. F. Mendon\c{c}a}\email{mendonca@physics.uq.edu.au}
\affiliation{Department of Physics, The University of Queensland, Queensland 4072, Australia}
\author{Reginaldo d. J. Napolitano}\email{reginald@ifsc.usp.br}
\affiliation{Instituto de F\'{\i}sica de S\~{a}o Carlos, Universidade de S\~{a}o Paulo, Caixa Postal 369,
             13560-970, S\~{a}o Carlos, SP, Brazil}
\author{Marcelo A. Marchiolli}\email{mamarchi@ift.unesp.br}
\affiliation{Instituto de F\'{\i}sica Te\'{o}rica, Universidade Estadual Paulista, Rua Pamplona 145, 01405-900,
             S\~{a}o Paulo, SP, Brazil}
\author{Christopher J. Foster}\email{foster@physics.uq.edu.au}
\affiliation{Department of Physics, The University of Queensland, Queensland 4072, Australia}
\author{Yeong-Cherng Liang}\email{ycliang@physics.usyd.edu.au}
\affiliation{School of Physics, The University of Sydney, NSW 2006, Australia}
\date{\today}
\begin{abstract}
We propose an alternative fidelity measure (namely, a measure of the degree of similarity) between quantum
states and benchmark it against a number of properties of the standard Uhlmann-Jozsa fidelity. This measure is a simple
function of the linear entropy and the Hilbert-Schmidt inner product between the given states and is thus, in comparison,
not as computationally demanding. It also features several remarkable properties such as being jointly concave and satisfying
all of {\it Jozsa's axioms}. The trade-off, however, is that it is supermultiplicative and does not behave monotonically
under quantum operations. In addition, new metrics for the space of density matrices are identified and the joint concavity
of the Uhlmann-Jozsa fidelity for qubit states is established.
\end{abstract}
\maketitle
\section{Introduction}\lb{sec:intro}

The understanding of the set of density matrices as a Riemannian manifold \cite{06Bengtsson} implies that a notion of
distance can be assigned to any pair of quantum states. In quantum information science, for instance, distance measures
between quantum states have proved to be useful resources in approaching a number of fundamental problems such as
quantifying entanglement \cite{97Vedral2275,98Vedral1619}, the design of optimized strategies for quantum control
\cite{07Branczyk012329,08Mendonca}, and quantum error correction \cite{05Reimpell080501,07Fletcher012338,06Reimpell,
06Kosut,08Kosut020502,07Yamamoto012327,05Yamamoto022322}. In addition, the concept of distinguishability between quantum
states \cite{95Fuchs} can be made mathematically rigorous and physically insightful thanks to the close relationship between
certain metrics for the space of density matrices and the error probability arising from various versions of the quantum
hypothesis-testing problem \cite{06Hayashi}. Distance measures are also regularly used in the laboratory to verify the
quality of the produced quantum states.

A widely used distance measure in the current literature (or, more precisely, a {\em fidelity measure} --- that is, a measure
of the degree of similarity --- between two general density matrices), is the so-called {\em Uhlmann-Jozsa fidelity}, $\F$.
Historically, this measure had its origins in the 1970s through a set of works by Uhlmann and Alberti \cite{76Uhlmann273,
83Alberti5,83Alberti25,83Alberti107}, who studied the problem of generalizing the quantum mechanical transition probability to
the broader context of $\ast$-algebras. The use of the term {\em fidelity} to designate Uhlmann's transition probability
formula is much more recent and initiated in the works of Schumacher \cite{95Schumacher2738} and Jozsa \cite{94Jozsa2315}.
Indeed, in an attempt to quantify the degree of similarity between a certain mixed state $\ro$ and a pure state $\ket{\psi}$,
Schumacher dubbed the transition probability $\bra{\psi} \ro \ket{\psi}$ the fidelity between the two states. In parallel,
Jozsa recognized Uhlmann's transition probability formula as a sensible extension of Schumacher's fidelity, where now the
measure of similarity is related to a pair of mixed states $\ro$ and $\sig$. Ever since, Uhlmann's transition probability
formula has been widely accepted as {\em the} generalization of Schumacher's fidelity.

The prevalence of this measure as one of the most used notions of distance in quantum information is not accidental, but
largely supported on a number of required and desired properties for the role. For example, $\F$ satisfies all of {\em Jozsa's
axioms}, that is, besides recovering Schumacher's fidelity in the case where one of the states is pure, the following three
additional properties also hold: First, $\F$ equals unity if and only if it is applied to two identical states; in other cases it lies between $0$ and $1$. Second, it is symmetric, i.e., the fidelity between $\ro$ and $\sig$ is the same as that between
$\sig$ and $\ro$. Third, it is invariant under any unitary transformation on the state space. Nevertheless, $\F$ is {\em not}
the unique measure satisfying these properties. A prominent alternative which also complies with Jozsa's axioms and shares
many other properties of $\F$, is given by the nonlogarithmic variety of the quantum Chernoff bound, $Q$, recently determined
in Ref.~\cite{07Audenaert160501}. In analogy with its classical counterpart \cite{00Nielsen}, the quantum Chernoff bound
determines --- in the limit of asymptotically many copies --- the minimum error probability incurred in discriminating between
two quantum states \cite{07Audenaert160501,fn:QCB2}.

Despite fulfilling the properties listed above, both $\F$ and $Q$ are, in general, unsatisfying measures from a practical
computational viewpoint. Although $\F$ can be expressed in a closed form in terms of $\ro$ and $\sig$, it involves successive
computation of the square roots of Hermitian matrices, which often compromises its use in analytical computations and
numerical experiments, especially when the fidelity measure must be computed many times. Even more serious is the case of $Q$,
which to date has only been defined variationally as the result of an optimization problem~\cite{fn:QCB2}. The question that
naturally arises is whether an easy-to-compute generalization of Schumacher's fidelity can be obtained. In this paper, we
provide a positive answer to this question and a thorough analysis of our proposed alternative fidelity measure $\Fn$.

Recently, we became aware of the very recent work of Miszczak {\em et al.}~\cite{08Miszczak} in
which $\Fn$ was introduced as an upper bound to the Uhlmann-Jozsa fidelity. In many ways our analysis of $\Fn$ is
complimentary to that provided in Ref. \cite{08Miszczak}; results in common are noted in the corresponding sections of our
paper.

Our paper is structured as follows. In order to provide a concrete ground for our proposal of $\Fn$ as an alternative fidelity
measure, we first reexamine, in Sec. \ref{sec:fid}, a set of basic properties of the Uhlmann-Jozsa fidelity. In Sec.
\ref{sec:new_meas} we formally introduce $\Fn$ and analyze it in the spirit of the properties reviewed in Sec. \ref{sec:fid}.
The computational efficiency of $\Fn$ is contrasted with a number of previously known distance or fidelity measures in Sec.
\ref{sec:computability}. We summarize our main results and discuss some possible avenues for future research in Sec.
\ref{sec:conclusion}.

\section{Uhlmann-Jozsa Fidelity}\lb{sec:fid}

In this section, we will briefly survey some physically appealing features inherent to the Uhlmann-Jozsa fidelity $\F$. In
Sec. \ref{sec:new_meas}, these features will be used as a reference for characterizing the proposed fidelity measure.

\subsection{Preliminaries}\lb{sec:fidprelim}

The Uhlmann-Jozsa fidelity $\F$ was originally introduced as a transition probability between two generic quantum states $\ro$ and $\sig$ \cite{76Uhlmann273}:
\be
\lb{eq:fid}
\F(\ro,\sig) \defeq \!\! \max_{\ket{\psi},\ket{\varphi}}{| \langle \psi | \varphi \rangle |^2} = \lbk \tr \lpar
\sqrt{\sqrt{\ro} \, \sig \sqrt{\ro}} \rpar \rbk^{2} .
\ee
Here, $\ket{\psi}$ and $\ket{\varphi}$ are restricted to be purifications of $\ro$ and $\sig$, while the second equality
indicates that the maximization procedure can be explicitly evaluated. At this stage, it is worth noting that it is not
uncommon to find $\sqrt{\F}$ being referred to, instead, as the fidelity (see, for example, Ref. \cite{00Nielsen}).

In Ref. \cite{94Jozsa2315}, Jozsa conjectured that Eq. \eqref{eq:fid} was the unique expression that satisfies a number of
natural properties expected for any generalized notion of fidelity \cite{fn:QCB}. Throughout, we shall refer to these as
{\em Jozsa's axioms}:
\begin{enumerate}
\item Normalization --- \ie, $\F(\ro,\sig) \in [0,1]$ with the upper bound attained iff $\ro = \sig$ (the {\em identity of
indiscernible} property).
\item Symmetry under swapping of the two states --- \ie, $\F(\ro,\sig) = \F(\sig,\ro)$.
\item Invariance under any unitary transformation ${\bf U}$ of the state space --- \ie, $\F({\bf U} \ro {\bf U}^{\dagger},
{\bf U} \sig {\bf U}^{\dagger}) = \F(\ro,\sig)$.
\item Consistency with Schumacher's fidelity when one of the states is pure --- \ie,
\be
\lb{eq:fid_schum}
\F(\ro,\ket{\psi} \! \bra{\psi}) = \bra{\psi} \ro \ket{\psi}
\ee
for arbitrary $\ro$ and $\ket{\psi}$.
\end{enumerate}

The proof that $\F$ satisfies all of Jozsa's axioms follows easily from the variational definition of Eq. \eqref{eq:fid}
(see, \eg, Ref. \cite{00Nielsen} for technical details). The remainder of this section discusses a number of less immediate
properties of $\F$.

\subsection{Concavity properties}\lb{sec:fidconc}

The concavity property of quantities like entropy, mutual information, and fidelity measure are often of theoretical interest in
the quantum information community \cite{00Nielsen}. In this regard, it is worth noting that a useful feature of $\F$ is its
{\em separate concavity} in each of its arguments; \ie, for $p_{1},p_{2}\geq 0$, $p_{1} + p_{2} = 1$, and arbitrary density
matrices $\ro_{1}$, $\ro_{2}$, $\sig_{1}$, and $\sig_{2}$, we have
\be
\lb{eq:sepconcF}
\F (p_{1} \ro_{1} + p_{2} \ro_{2},\sig_{1}) \geq p_{1} \F(\ro_{1},\sig_{1}) + p_{2} \F(\ro_{2},\sig_{1}) .
\ee
By symmetry, concavity in the second argument follows from Eq. \eqref{eq:sepconcF}. Separate concavity can be proven
\cite{76Uhlmann273,94Jozsa2315} using the variational definition of $\F$ from Eq. \eqref{eq:fid}.

While it is known that $\sqrt{\F}$ is {\em jointly concave} \cite{83Alberti5,00Uhlmann407}, \ie,
\begin{gather}
\lb{eq:jconcsqrtF}
\sqrt{\F} (p_{1} \ro_{1} + p_{2} \ro_{2},p_{1} \sig_{1} + p_{2} \sig_{2}) \nn \\
\geq p_{1} \sqrt{\F}(\ro_{1},\sig_{1}) + p_{2} \sqrt{\F}(\ro_{2},\sig_{2}) ,
\end{gather}
it is also known that the Uhlmann-Jozsa fidelity $\F$ does {\em not}, in general, share the same enhanced concavity property
\cite{fn:Concavity}.

\subsection{Multiplicativity under tensor products}\lb{sec:fidmultiplic}

Another neat mathematical property of $\F(\ro,\sig)$ is that it is multiplicative under tensor products: for any density matrices
$\ro_{1}$, $\ro_{2}$, $\sig_{1}$ and $\sig_{2}$:
\be
\F(\ro_{1} \otimes \ro_{2},\sig_{1} \otimes \sig_{2}) = \F(\ro_{1},\sig_{1}) \F(\ro_{2},\sig_{2}) .
\ee
This identity follows easily from the following facts: for any Hermitian matrices $\mathds{A}$ and $\mathds{B}$, (i) $\tr(\mathds{A}
\otimes \mathds{B}) = \tr(\mathds{A}) \, \tr(\mathds{B})$ and (ii) $\sqrt{\mathds{A} \otimes \mathds{B}} = \sqrt{\mathds{A}}
\otimes \sqrt{\mathds{B}}$.

An immediate consequence of this result is that for two physical systems, described by $\ro$ and $\sig$, a measure of their degree
of similarity given by $\F$ remains unchanged even after appending each of them with an uncorrelated ancillary state $\btau$ ---
\ie, $\F(\ro \otimes \btau,\sig \otimes \btau) = \F(\ro,\sig)$.

\subsection{Monotonicity under quantum operations}\lb{sec:fidmonotone}

Given that $\F(\ro,\sig)$ serves as a kind of measure for the degree of similarity between two quantum states $\ro$ and $\sig$,
one might expect that a general quantum operation $\E$ will make them less distinguishable and, hence, more similar according to
$\F$ \cite{00Nielsen}:
\be
\lb{eq:Dmonotone}
\F(\E(\ro),\E(\sig)) \geq \F(\ro,\sig) .
\ee
Indeed, it is now well known that Eq. \eqref{eq:Dmonotone} holds true \cite{83Alberti107} for an arbitrary quantum operation
described by a completely positive trace-preserving (CPTP) map $\E : \ro \mapsto \E(\ro)$. Inequality \eqref{eq:Dmonotone}
qualifies $\F$ as a {\em monotonically increasing measure} under CPTP maps and can be considered the quantum analog of the
classical {\em information-processing inequality} --- which expresses that the amount of information should not increase via
any information processing.

On a related note, it is worth noting that any measure $\mathcal{M}$ which is (i) unitarily invariant, (ii) jointly concave
(convex), and (iii) invariant under the addition of an ancillary system is also monotonically increasing (decreasing) under CPTP
maps \cite{fn:Monotonicity}. Clearly, since $\sqrt{\F}$ satisfies all the above-mentioned conditions, Eq. \eqref{eq:Dmonotone}
also follows by simply squaring the corresponding monotonicity inequality for $\sqrt{\F}$.

\subsection{Related metrics}\label{sec:fidmetric}

The Uhlmann-Jozsa fidelity by itself is not a metric (for a quick review of metrics, see Appendix \ref{app:metrics}). However,
one may well expect that a metric, which is a measure of distance, can be built up from a measure of similarity such as $\F$.
Indeed, the functionals
\begin{align}
\lb{eq:AF}
A[\F(\ro,\sig)] & \defeq \arccos \lpar {\sqrt{\F(\ro,\sig)}} \rpar , \\
\lb{eq:BF}
B[\F(\ro,\sig)] & \defeq \sqrt{2 - 2 \sqrt{\F(\ro,\sig)}} , \\
\lb{eq:CF}
C[\F(\ro,\sig)] & \defeq \sqrt{1-\F(\ro,\sig)}
\end{align}
exhibit such metric properties (see Refs. \cite{95Uhlmann461,00Nielsen,69Bures199,92Hubner239,05Gilchrist062310,06Rastegin} and
also Appendix \ref{app:metric_C} for more details). In particular, these functionals are now commonly known in the literature,
respectively, as the {\em Bures angle} \cite{00Nielsen}, the {\em Bures distance} \cite{69Bures199,92Hubner239}, and the
{\em sine distance} \cite{06Rastegin}.

\subsection{Trace distance bounds}\lb{sec:fidtrdistbnd}

An important distance measure in quantum information is the metric induced by  the trace norm $\| \cdot \|_{\rm tr}$, which is
commonly referred to as the trace distance \cite{00Nielsen}:
\be
\lb{eq:trdist}
\D(\ro,\sig) = \tfrac{1}{2} \| \ro - \sig \|_{\rm tr} .
\ee
The trace distance is an exceedingly successful distance measure: it is a metric (as is any distance induced by norms),
unitarily invariant \cite{97Bhatia}, jointly convex \cite{00Nielsen}, decreases under CPTP maps \cite{94Ruskai1147}, and in
the qubit case, is proportional to the Euclidean distance between the Bloch vectors in the Bloch ball. The trace distance is
also closely related to the minimal probability of error on attempts to distinguish between a single copy of two
nonorthogonal quantum states \cite{76Helstrom}. For all of these reasons, one is generally interested to determine how other
distance measures relate with the trace distance.

The following functions of the Uhlmann-Jozsa fidelity were shown in Ref.~\cite{99Fuchs1216} to provide tight bounds for $\D$
\cite{fn:TrD}:
\be
\lb{eq:fuchs_ineq}
1 - \sqrt{\F(\ro,\sig)} \leq \D(\ro,\sig) \leq \sqrt{1 - \F(\ro,\sig)} .
\ee
In fact, the stronger lower bound $1 - \F \leq \D$ holds if $\ro$ and $\sig$ have support on a common two-dimensional Hilbert
space \cite{01Spekkens012310} (\eg, any pair of qubit states) or if at least one of the states is pure \cite{00Nielsen}.

From these inequalities, one can conclude a type of qualitative equivalence between the Uhlmann-Jozsa fidelity $\F$ and the
trace distance $\D$: whenever $\F$ is small, $\D$ is large and whenever $\F$ is large, $\D$ is small.

\section{Alternative fidelity measure}\lb{sec:new_meas}
\subsection{Preliminaries}

We shall now turn attention to our proposed measure of the degree of similarity between two quantum states $\ro$ and $\sig$ --- namely,
\be
\lb{eq:fidnew}
\Fn(\ro,\sig) = \tr (\ro \sig) + \sqrt{1 - \tr(\ro^{2})} \sqrt{1 - \tr(\sig^{2})} .
\ee
This is simply a sum of the Hilbert-Schmidt inner product between $\ro$ and $\sig$ and the geometric mean between their linear
entropies. It is worth noting that the same quantity --- by the name {\em superfidelity} --- has been independently
introduced in Ref. \cite{08Miszczak} as an upper bound for $\F$.

Remarkably, when applied to qubit states, $\Fn$ is precisely the same as $\F$. This observation follows easily from the fact that
for density matrices of dimension $d = 2$, it is valid to write
\be
\Fn(\ro,\sig) \Bigr|_{d=2} = \tr (\ro \sig) + 2 \sqrt{\det{(\ro)}} \sqrt{\det{(\sig)}} ,
\ee
which is just an alternative expression of $\F$ for qubit states \cite{93Hubner226,92Hubner239}.

When $d > 2$, however, $\Fn$ no longer recovers $\F$, but can be seen as a simplified version of the fidelity measure proposed by
Chen and collaborators \cite{02Chen054304}, which reads as
\be
\F_{\mbox{\tiny $C$}}(\ro,\sig) = \frac{1-r}{2} + \frac{1+r}{2} \, \Fn(\ro,\sig) ,
\ee
where $r = 1/(d-1)$ and $d$ is the dimension of the state space of $\ro$ and $\sig$. Moreover, it is straightforward to verify
that while $\Fn$ reduces to the Schumacher's fidelity [the right-hand side of Eq. \eqref{eq:fid_schum}] when one of the states is pure; the
same cannot be said for $\F_{\mbox{\tiny $C$}}$.

It is not difficult to see from Eq. \eqref{eq:fidnew} that $\Fn$ satisfies Jozsa's axioms 2, 3, and 4 as enumerated in Sec.
\ref{sec:fidprelim}. The non-negativity of $\Fn$ required by axiom 1 is also immediate from the definition. As a result, $\Fn$
is an acceptable generalization of Schumacher's fidelity according to Jozsa's axioms if the following proposition is true.
\begin{proposition}\lb{prop:jozsa}
$\Fn(\ro,\sig) \leq 1$ holds for arbitrary density matrices $\ro$ and $\sig$, with saturation if and only if $\ro = \sig$.
\end{proposition}
\begin{proof}
To begin with, recall that any $d \times d$ density matrix can be expanded in terms of an orthonormal basis of Hermitian matrices
$\{ v_{k} \}_{k=0}^{d^{2}-1}$ such that $\tr(v_{i} v_{j}) = \delta_{ij}$ (see, for example, Refs. \cite{03Byrd062322,03Kimura339}).
In particular, if we let $\vec{\Upsilon} \defeq (v_{0}, \ldots, v_{d^{2}-1})$, then $\ro$ and $\sig$ admit the following
decomposition:
\be
\lb{eq:param}
\ro = \vec{r} \cdot \vec{\Upsilon} \quad \mbox{and} \quad \sig = \vec{s} \cdot \vec{\Upsilon} ,
\ee
where $\vec{r}$ and $\vec{s}$ are real vectors with $d^{2}$ entries (corresponding to the expansion coefficients which can be
determined using the orthonormality condition). Since $\ro$ and $\sig$ are density matrices, $\vec{r}$ and $\vec{s}$ satisfy
$0 \leq \vec{r} \cdot \vec{s}\leq 1$ and $ r,s \leq 1$, where $r = \| \vec{r} \|$ and $s = \| \vec{s} \|$.

Using the expansion of Eq. \eqref{eq:param} in Eq. \eqref{eq:fidnew}, we arrive at the following alternative expression of $\Fn$,
\begin{align}
\lb{eq:fN}
f_{\mbox{\tiny $N$}}(\vec{r},\vec{s}) & = \vec{r} \cdot \vec{s} + \sqrt{1-r^{2}} \sqrt{1-s^{2}} \\
& = \vec{R} \cdot \vec{S} ,
\end{align}
where, in the second line, we have defined two {\em unit} vectors in $\mathbb{R}^{d^{2}+1}$, explicitly,
\be
\vec{R} \defeq \lpar \vec{r}, \sqrt{1-r^{2}} \rpar \quad \mbox{and} \quad \vec{S} \defeq \lpar \vec{s}, \sqrt{1-s^{2}} \rpar .
\ee
The normalization of $\vec{R}$ and $\vec{S}$ implies that $\Fn(\ro,\sig) = \vec{R} \cdot \vec{S} \leq 1$, with saturation
if and only if $\vec{R} = \vec{S}$, or equivalently $\ro = \sig$.
\end{proof}

\subsection{Concavity properties}\lb{sec:fnconc}

As with $\sqrt{\F}$, the measure $\Fn$ is jointly concave in its two arguments; \ie, for $ p_{1},p_{2} \geq 0$, $p_{1} + p_{2}
= 1$, and arbitrary density matrices $\ro_{1}$, $\ro_{2}$, $\sig_{1}$, and $\sig_{2}$, we have
\brr
\lb{eq:fnjconc}
& \Fn(p_{1} \ro_{1} + p_{2} \ro_{2},p_{1} \sig_{1} + p_{2} \sig_{2}) \geq \nn \\
& p_{1} \Fn(\ro_{1},\sig_{1}) + p_{2} \Fn(\ro_{2},\sig_{2}) .
\err
Since $\F$ fails to be jointly concave in general, $\Fn$ has a stronger concavity property. Remarkably, given the equivalence between
$\F$ and $\Fn$ in the $d = 2$ case, the result of this section implies that $\F$ is jointly concave when restricted to qubit states.

The rest of this section concerns a proof of this concavity property of $\Fn$. We start by proving the following lemma, which
provides a useful alternative expression of inequality \eqref{eq:fnjconc}.

\begin{lemma}\lb{lemmaconc}
Define a function $F:[0,1] \to \mathbb{R}$ by
\begin{multline}
\lb{eq:Fx}
F(x) \defeq (\vec{r} + x \vec{u}) \cdot (\vec{s} + x \vec{v}) \\
 + \sqrt{1 - \| \vec{r} + x \vec{u} \|^{2}} \sqrt{1 - \| \vec{s} + x \vec{v} \|^{2}} .
\end{multline}
Given the density matrices $\ro_{1}$, $\ro_{2}$, $\sig_{1}$, and $\sig_{2}$, there exist vectors $\vec{r}$, $\vec{s}$, $\vec{u}$, $\vec{v}
\in \mathbb{R}^{d^{2}}$ and $x \in [0,1]$ such that the inequality
\be
\lb{eq:linebelow}
F(x) \geq (1-x)F(0) + x F(1)
\ee
is equivalent to Eq. \eqref{eq:fnjconc}.
\end{lemma}
\begin{proof}
The proof is by construction. Using the parametrization of Eq. (\ref{eq:param}) for the density matrices in inequality
\eqref{eq:fnjconc}, we obtain the following equivalent inequality for the vectors $\vec{r}_{i}$ and $\vec{s}_{i}$:
\brr
\lb{eq:ineqfn}
& f_{\mbox{\tiny $N$}}(p_{1} \vec{r}_{1} + p_{2} \vec{r}_{2},p_{1} \vec{s}_{1} + p_{2} \vec{s}_{2}) \geq \nn \\
& p_{1} f_{\mbox{\tiny $N$}} (\vec{r}_{1},\vec{s}_{1}) + p_{2} f_{\mbox{\tiny $N$}} (\vec{r}_{2},\vec{s}_{2}) ,
\err
where the function $f_{\mbox{\tiny $N$}}$ was defined in Eq. \eqref{eq:fN}.

A straightforward computation shows that inequality \eqref{eq:linebelow} is identical to inequality \eqref{eq:ineqfn} when
we identify $x \equiv p_{2}$, $1 - x \equiv p_{1}$, and set
\be
\lb{eq:rusv}
\begin{array}{rclcrcl}
        \vec{r} &=& \vec{r}_{1} , \quad \vec{u} = \vec{r}_{2} - \vec{r}_{1} , \\
        \vec{s} &=& \vec{s}_{1} , \quad \vec{v} = \vec{s}_{2} - \vec{s}_{1} .
    \end{array}
\ee
\end{proof}

If $F(x)$ has negative concavity in $x \in [0,1]$, then inequality \eqref{eq:linebelow} is automatically satisfied as it
establishes that the straight line connecting the points $(0,F(0))$ and $(1,F(1))$ lies below the curve $\{(x,F(x)) | x \in
[0,1] \}$. As a result, the joint concavity of $\Fn$ is proved with the following proposition.
\begin{proposition}\lb{prop:concFx}
For $x \in [0,1]$, and $\vec{r}$, $\vec{s}$, $\vec{u}$, $\vec{v}\in \mathbb{R}^{d^{2}}$ specified in Eq. (\ref{eq:rusv}), the
function $F(x)$ [cf. Eq. (\ref{eq:Fx})] satisfies
\be
\frac{d^{2} F(x)}{dx^{2}} \leq 0
\ee
and hence $\Fn$ is jointly concave.
\end{proposition}
\noindent The proof of this proposition is given in Appendix \ref{app:concFx}.

\subsection{Multiplicativity under tensor product} \lb{sec:fnmultipl}

In contrast with $\F$, the new measure $\Fn$ is not multiplicative under tensor products. In fact, it is generally not even invariant
under the addition of an uncorrelated ancilla prepared in the state $\btau$. In this case, $\Fn$ between the resulting states reads
as
\begin{align*}
& \Fn(\ro \otimes \btau,\sig \otimes \btau) = \tr (\ro \sig) \tr (\btau^{2}) \nn \\
& \quad + \sqrt{1 - \tr (\ro^{2}) \tr (\btau^{2})} \sqrt{1 - \tr (\sig^{2}) \tr (\btau^{2})} ,
\end{align*}
where the left-hand side equals $\Fn(\ro,\sig)$ if and only if $\tr (\btau^{2}) = 1$ or, in other words, if and only if $\btau$ is a pure state. More generally, it
can be shown that $\Fn$ is supermultiplicative, \ie,
\be
\Fn(\ro_{1} \otimes \ro_{2},\sig_{1} \otimes \sig_{2}) \geq \Fn(\ro_{1},\sig_{1}) \Fn(\ro_{2},\sig_{2}) .
\ee
A proof of this property is given in Appendix \ref{app:super-multplicativity}; a similar proof was independently obtained in
Ref. \cite{08Miszczak}.

\subsection{Monotonicity under quantum operations}\lb{sec:fnmonotone}

That $\Fn$ is only supermultiplicative may be a first sign that it may not behave monotonically under CPTP maps. In fact, as we
shall see below, Ozawa's counterexample \cite{00Ozawa158} to the claimed monotonicity of the Hilbert-Schmidt distance \cite{99Witte14}
can also be used to show that $\Fn$  does not behave monotonically under CPTP maps.

Let $\widetilde{\ro}$ and $\widetilde{\sig}$ be two two-qubit density matrices, written in the product basis as
\be
\lb{eq:Ozawa}
    \widetilde{\ro} = \tfrac{1}{2} \lpar
    \begin{array}{cccc}
    1 & 0 & 0 & 0 \\
    0 & 1 & 0 & 0 \\
    0 & 0 & 0 & 0 \\
    0 & 0 & 0 & 0
    \end{array}
    \rpar \quad \mbox{and} \quad
    \widetilde{\sig} = \tfrac{1}{2} \lpar
    \begin{array}{cccc}
    0 & 0 & 0 & 0 \\
    0 & 0 & 0 & 0 \\
    0 & 0 & 1 & 0 \\
    0 & 0 & 0 & 1
    \end{array} \rpar ,
\ee
and consider the (trace-preserving) quantum operations of tracing over the first or second qubit. A straightforward computation
shows that if the first qubit is traced over, then
\be
\lb{eq:monot1}
\Fn(\tr_{1}(\widetilde{\ro}),\tr_{1}(\widetilde{\sig})) = 1 > \tfrac{1}{2} = \Fn(\widetilde{\ro},\widetilde{\sig}) ,
\ee
which satisfies the desired monotonicity property. However, if instead the second subsystem is discarded, we find
\be
\lb{eq:monot2}
\Fn(\tr_{2}(\widetilde{\ro}),\tr_{2}(\widetilde{\sig})) = 0 < \tfrac{1}{2} = \Fn(\widetilde{\ro},\widetilde{\sig}) .
\ee
Together, Eqs. \eqref{eq:monot1} and \eqref{eq:monot2} show that $\Fn$ is neither monotonically increasing nor decreasing under
general CPTP maps.

A natural question that follows is whether $\Fn$ features a weaker form of monotonicity. For example, do arbitrary projective
measurements --- with the measurement outcomes forgotten --- give rise to a higher value of $\Fn$ for the resulting pair of
states? An affirmative answer would follow from a proof of the inequality
\be
\lb{eq:pinching}
\Fn \lpar \sum_{i} {\bf P}_{i} \ro {\bf P}_{i}, \sum_{i} {\bf P}_{i} \sig {\bf P}_{i} \rpar \geq \Fn(\ro,\sig)
\ee
for any complete set of orthonormal projectors ${\bf P}_{i}$ and for arbitrary density matrices $\ro$ and $\sig$.

It is a simple  exercise to prove Eq. \eqref{eq:pinching} for the particular case where either of the commutation rules
$[ {\bf P}_{i},\ro ] = 0$ or $[ {\bf P}_{i},\sig ] = 0$ is observed for all values of $i$. Whether the same conclusion can be
drawn for the more general, noncommutative cases remains to be seen. In this regard, we note that a preliminary numerical
search favors the validity of Eq. \eqref{eq:pinching}.

\subsection{Related metrics}\lb{sec:fnmetric}

In parallel to the metrics $A[\F]$, $B[\F]$, and $C[\F]$ introduced in Sec. \ref{sec:fidmetric}, we define
\begin{align}
\lb{eq:AFn}
A[\Fn(\ro,\sig)] & \defeq \arccos \lpar \sqrt{\Fn(\ro,\sig)} \rpar , \\
\lb{eq:BFn}
B[\Fn(\ro,\sig)] & \defeq \sqrt{2 - 2 \sqrt{\Fn(\ro,\sig)}} , \\
\lb{eq:CFn}
C[\Fn(\ro,\sig)] & \defeq \sqrt{1 - \Fn(\ro,\sig)} ,
\end{align}
and prove that while $C[\Fn]$ preserves the metric properties, both $A[\Fn]$ and $B[\Fn]$ {\em do not} always obey the triangle
inequality
\be
\lb{eq:triang}
X[\Fn(\ro,\sig)] \leq X[\Fn(\ro,\btau)] + X[\Fn(\btau,\sig)] ,
\ee
where $X$ here refers to either $A$, $B$, or $C$. For example, consider the qutrit density matrices, $\ro = \openone_{3}/3$,
\be
\lb{eq:example_states}
\sig = \lpar \begin{array}{ccc}
       1 & 0 & 0 \\
       0 & 0 & 0 \\
       0 & 0 & 0 \end{array} \rpar \mbox{and} \;
\btau = \lpar \begin{array}{ccc}
       0.90 & 0.04 & 0.03 \\
       0.04 & 0.05 & 0.02 \\
       0.03 & 0.02 & 0.05 \end{array} \rpar .
\ee

\begin{table}[h!]
\centering
\caption{A numerical test of the triangle inequality for $A[\Fn]$, $B[\Fn]$, and $C[\Fn]$.}
\begin{ruledtabular}
\begin{tabular}{r|cc}
$X$ & $X[\Fn(\ro,\sig)]$ & $X[\Fn(\ro,\btau)] + X[\Fn(\btau,\sig)]$ \\
\hline
$A$ & $0.9553$ & $0.9241$ \\
$B$ & $0.9194$ & $0.9137$ \\
$C$ & $0.8165$ & $0.8828$ \\
\end{tabular} \lb{table:numtest}
\end{ruledtabular}
\end{table}
Numerical computation of the quantities appearing in the triangle inequality gives rise to Table \ref{table:numtest}.  Note that
for $X = A,B$, the first column dominates the second; \ie, the triangle inequality is violated and therefore neither $A[\Fn]$
nor $B[\Fn]$ is a metric. For $X = C$, no violation is observed for the above density matrices. Next, we prove that this is the
case for any three density matrices $\ro$, $\sig$ and $\btau$; thus, $C[\Fn]$ is a metric.
\begin{proposition}
\lb{Pro:metric:C(Fn)}
The quantity $C[\Fn(\ro,\sig)]$ is a metric for the space of density matrices.
\end{proposition}

To prove this proposition, we will make use of the following theorem due to Schoenberg \cite{38Schoenberg522} (see also
\cite[Chap. 3, Proposition 3.2]{84Berg}). We state here an abbreviated form of the theorem sufficient for our present purposes.
\begin{theorem}[Schoenberg]
\lb{thm:schoenberg}
Let $\mathcal{X}$ be a nonempty set and $K : \mathcal{X} \times \mathcal{X} \to \mathbb{R}$ a function such that $K(x,y) = K(y,x)$
and $K(x,y) \geq 0$ with saturation iff $x = y$, for all $x,y \in \mathcal{X}$. If the implication
\be
\lb{eq:ndk}
\sum_{i = 1}^{n} c_{i} = 0 \Rightarrow \sum_{i,j = 1}^{n} K(x_{i},x_{j}) \, c_{i} c_{j} \leq 0
\ee
holds for all $n \geq 2$, $\{ x_{1}, \ldots, x_{n} \} \subseteq \mathcal{X}$, and $\{ c_{1}, \ldots, c_{n} \} \subseteq \mathbb{R}$,
then $\sqrt{K}$ is a metric.
\end{theorem}

We make a small digression at this point to remark that, in spite of its successful application on the grounds of classical
probability distance measures \cite{00Topsoe1602,03Topsoe,04Fuglede}, Schoenberg's theorem has received almost no attention by the
quantum information community. In this paper, besides proving the metric properties of $C[\Fn]$, we will also make use of Schoenberg's
theorem to provide independent proofs of the metric properties of $B[\F(\ro,\sig)]$ and $C[\F(\ro,\sig)]$ (see Appendix
\ref{app:metric_C}).

\begin{proof}[Proof of Proposition \ref{Pro:metric:C(Fn)}]
Clearly, from the definition of $C^{2}[\Fn(\ro,\sig)]$, it is easy to see that it inherits from $\Fn(\ro,\sig)$ the property of
being symmetric in its two arguments and that $C^{2}[\Fn(\ro,\sig)] \geq 0$ with saturation iff $\ro = \sig$. So, to apply Theorem
\ref{thm:schoenberg}, we just have to show that for any set of density matrices $\{ \ro_{i} \}_{i=1}^{n}$ $(n \geq 2)$ and real
numbers $\{ c_{i} \}_{i=1}^{n}$ such that $\sum_{i=1}^{n} c_{i} = 0$, it is true that
\be
\sum_{i,j=1}^{n} C^{2}[\Fn(\ro_{i},\ro_{j})] \, c_{i} c_{j} \leq 0 .
\ee
This follows straightforwardly by exploiting the zero-sum property of the (real) coefficients $c_{i}$ and the linearity of the trace,
\begin{align}
& \sum_{i,j=1}^{n} \lbk 1 - \tr (\ro_{i} \ro_{j}) - \sqrt{1 - \tr (\ro_{i}^{2})} \sqrt{1 - \tr (\ro_{j}^{2})} \rbk c_{i} c_{j} \nn \\
= & - \tr \lbk \lpar \sum_{i=1}^{n} c_{i} \ro_{i} \rpar^{2} \rbk - \lbk \sum_{i=1}^{n} c_{i} \sqrt{1 - \tr (\ro_{i}^{2})} \rbk^{2}
\leq 0 ,
\end{align}
which concludes the proof.
\end{proof}

We note that a proof of the metric property of $\sqrt{2} C[\Fn(\ro,\sig)]$ --- by the name {\em modified Bures distance} ---
was independently provided by Ref. \cite{08Miszczak}. The proof provided above is significantly shorter thanks to the power of
Schoenberg's theorem.

\subsection{Trace distance bounds}\lb{sec:fntrdistbnd}

In Sec. \ref{sec:fidtrdistbnd}, we have seen that a kind of qualitative equivalence between $\D$ and $\F$ can be established through
the bounds on $\D$ given by functions of $\F$; cf. Eq. \eqref{eq:fuchs_ineq}. Here, we will provide similar bounds on $\D$ in terms
of functions of $\Fn$.

\begin{proposition}
\lb{prob:lb}
For any two density matrices $\ro$ and $\sig$ of dimension $d$, the trace distance $\D(\ro,\sig)$ satisfies the following upper bound:
\be
\lb{eq:upbndFn}
\D(\ro,\sig) \leq \sqrt{\frac{\mathfrak{r}}{2}} \sqrt{1 - \Fn(\ro,\sig)} ,
\ee
where $\mathfrak{r}\defeq {\rm rank}(\ro - \sig)$. Moreover, this upper bound on $\D$ can be saturated with states of the form
\be
\lb{eq:satupper}
\ro = \frac{{\bf U} {\rm diag} [ \bLamb_{d} ] {\bf U}^{\dagger}}{\tr \lpar {\rm diag} [ \bLamb_{d} ] \rpar} \; \; \mbox{and} \; \;
\sig = \frac{{\bf U} {\rm diag} [ P(\bLamb_{d}) ] {\bf U}^{\dagger}}{\tr \lpar {\rm diag} [ \bLamb_{d} ] \rpar} ,
\ee
where ${\bf U}$ is an arbitrary unitary matrix of dimension $d$, $\bLamb_{d}$ is an ordered list of $d$ elements taking values in the
set $\{ \lambda_{1},\lambda_{2} \}$ ($\lambda_{1}, \lambda_{2} \geq 0$, but not simultaneously zero) and $P(\bLamb_{d})$ is the list
formed by some permutation of the elements in $\bLamb_{d}$.
\end{proposition}

\begin{proof}
Note that the product of square roots in the expression of $\Fn$, Eq. \eqref{eq:fidnew}, is the geometric mean between the linear
entropies of $\ro$ and $\sig$. It then follows from the inequality of arithmetic and geometric means that
\be
\lb{eq:AMGM}
\frac{1 - \tr (\ro^{2})}{2} + \frac{1 - \tr (\sig^{2})}{2} \geq \sqrt{1 - \tr (\ro^{2})} \sqrt{1 - \tr (\sig^{2})} ,
\ee
which can be reexpressed as the following inequality after summation of $\tr (\ro \sig)$ to both sides:
\be
\lb{eq:HSFn}
\| \ro - \sig \|_{\idhs} \leq \sqrt{2 \lbk 1 - \Fn(\ro,\sig) \rbk} .
\ee
Here, $\| {\bf X} \|_{\idhs} \defeq \sqrt{\tr ({\bf X}^{\dagger} {\bf X})}$ is the Hilbert-Schmidt norm (also known as Frobenius norm),
defined for an arbitrary matrix ${\bf X}$. The Hilbert-Schmidt norm and the trace norm $\| {\bf X} \|_{\rm tr} \defeq \tr \lpar
\sqrt{{\bf X}^{\dagger} {\bf X}} \rpar$ are related according to \cite{fn:Tr-HS}
\be
\lb{eq:normtrhs}
\| {\bf X} \|_{\rm tr} \leq \sqrt{\mathfrak{x}} \| {\bf X} \|_{\idhs} ,
\ee
where $\mathfrak{x} \defeq {\rm rank} ({\bf X})$. Used in Eq. \eqref{eq:HSFn}, the above inequality leads to the desired result
\be
\D(\ro,\sig) = \tfrac{1}{2} \| \ro - \sig \|_{\rm tr} \leq \sqrt{\frac{\mathfrak{r}}{2}} \sqrt{1 - \Fn(\ro,\sig)} .
\ee

To prove that the states in Eq. \eqref{eq:satupper} saturate this bound, we first note that because those states are isospectral,
their linear entropies are identical and hence inequality \eqref{eq:AMGM} is saturated. To prove saturation of inequality
\eqref{eq:normtrhs}, simply use Eq. \eqref{eq:satupper} to compute
\begin{align}
\!\!\!\!\!\! & \| \ro - \sig \|_{\rm tr} = \tr \lbk \sqrt{(\ro - \sig)^{2}} \rbk = \frac{\mathfrak{r} | \lambda_{1} -
\lambda_{2} |}{\tr ({\rm diag} [\bLamb_{d}])} , \\
\!\!\!\!\!\! & \| \ro - \sig \|_{\idhs} = \sqrt{\tr \lbk (\ro - \sig)^{2} \rbk} = \frac{\sqrt{\mathfrak{r}} | \lambda_{1} -
\lambda_{2} |}{\tr ({\rm diag} [\bLamb_{d}])} ,
\end{align}
from which the identity $\| \ro - \sig \|_{\rm tr} = \sqrt{\mathfrak{r}} \| \ro - \sig \|_{\idhs}$ is immediate.
\end{proof}

\begin{figure}[!t]
\flushleft(a)\\\centering
\includegraphics[width=8cm]{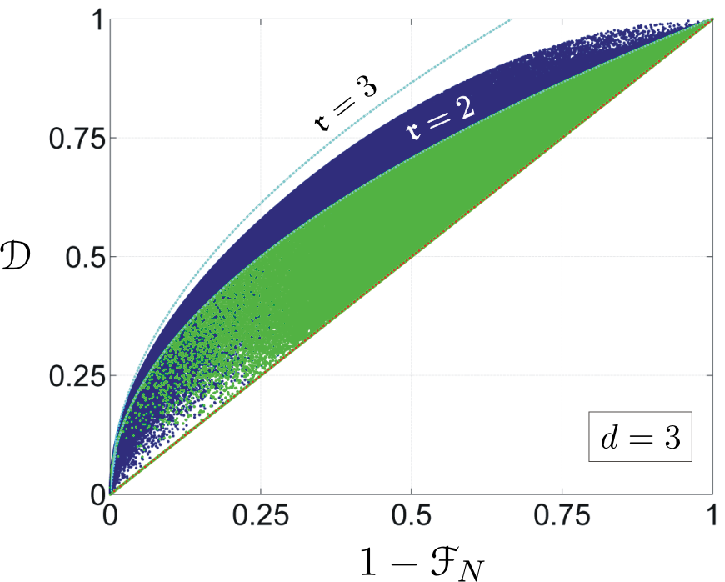}
\flushleft(b)\\ \centering
\includegraphics[width=8cm]{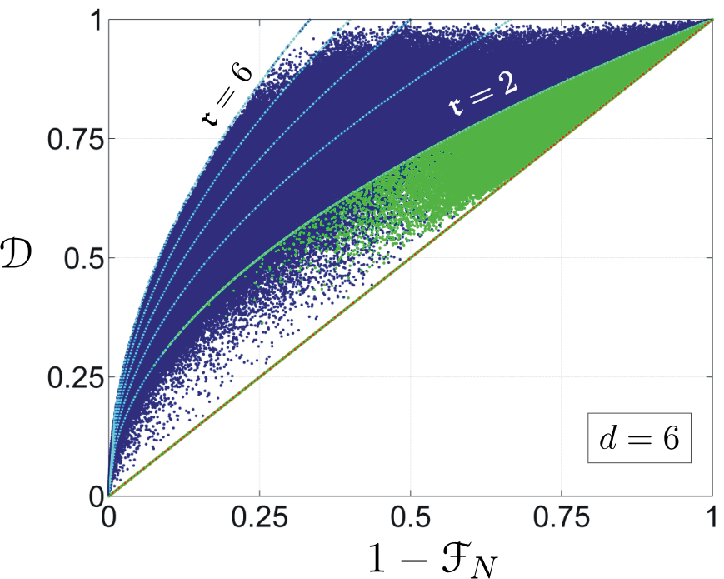}
\caption{(Color online) Plot of the trace distance $\D(\ro,\sig)$ versus $1 - \Fn(\ro,\sig)$ for $4 \times 10^{6}$ pairs of randomly
generated $\ro$ and $\sig$ with (a) $d=3$ and (b) $d=6$. The darker (blue) points are generated using pairs of mixed states whereas
the lighter (green) points are generated using at least one pure state. The antidiagonal solid line is the conjectured lower bound,
whereas the upper bounds given by Eq. \eqref{eq:upbndFn} are represented by the dashed curves (cyan) --- one for each integer value of
$\mathfrak{r} \in [2,d]$. A gap can be clearly noticed in plot (a) between the distribution of states and the {\em absolute upper
bound} --- \ie, the right-hand side of inequality \eqref{eq:upbndFn} with $\mathfrak{r} = d$. Such a gap occurs whenever $d$ is odd. However, no gap
is observed in plot (b) between the bulk of randomly generated states and the {\em absolute upper bound}. In fact, this bound can
be saturated by density matrices of the form given by Eq. \eqref{eq:satupper} whenever $d$ is even.}
\end{figure}
How good are  these upper bounds? With some thought, it is not difficult to conclude that the states arising from Eq.
\eqref{eq:satupper} can only have even $\mathfrak{r}$ and are thus unable to saturate the upper bound of Eq.
\eqref{eq:upbndFn} for odd $\mathfrak{r}$. Nonetheless, from our numerical studies, it seems like the {\em absolute upper
bound} --- corresponding to the choice $\mathfrak{r} = d$ on the right-hand side of Eq. \eqref{eq:upbndFn} --- is actually unachievable by
{\em any} states if $d$ is odd. An illustration of this peculiarity can be seen in Fig. 1(a), where the upper bound
corresponding to $\mathfrak{r} = 3$ is well separated from the region attainable by physical states. In contrast, for every
even $d$, the states given by Eq. \eqref{eq:satupper} do trace out a tight boundary for the region attainable with physical
states, as shown in Fig. 1(b) for $d=6$.

On the other hand, it can also be seen from Fig. 1 that no points occur in the region where $\D \leq 1 - \Fn$. Indeed,
intensive numerical studies for $d = 3, 4, \ldots, 50$ have not revealed a single pair of density matrices which contributed
to a point in this region. This suggests that the following lower bound on $\D$, in terms of $\Fn$, may well be established.
\begin{conjecture}
The trace distance $\D(\ro,\sig)$ and the measure $\Fn(\ro,\sig)$ between two quantum states $\ro$ and $\sig$ satisfy
\be
\D(\ro,\sig) \ge 1 - \Fn(\ro,\sig) .
\ee
\end{conjecture}
\noindent In relation to this, it is also worth noting that the following (weaker) lower bound can readily be established via a
recent result given in Ref. \cite{08Miszczak}:
\begin{proposition}
The trace distance $\D(\ro,\sig)$ and the measure $\Fn(\ro,\sig)$ between two quantum states $\ro$ and $\sig$ satisfy the
inequality
\be
\D(\ro,\sig) \ge 1 - \sqrt{\Fn}(\ro,\sig) .
\ee
\end{proposition}
\begin{proof}
This lower bound on $\D$ follows immediately from the lower bound on $\D$ given in inequality \eqref{eq:fuchs_ineq} and the
inequality $\F \leq \Fn$ recently established in Ref. \cite{08Miszczak}.
\end{proof}

As with the Uhlmann-Jozsa fidelity $\F$, we can thus infer that whenever $\Fn$ is large enough, $\D$ is close to zero and
whenever $\Fn$ is close to zero, $\D$ is close to unity. However --- as should be clear from Fig. 1(b) --- the converse
implication is not necessarily true.

\section{Computational Efficiency}\lb{sec:computability}

For two general density matrices $\ro$ and $\sig$, analytical evaluation of the Uhlmann-Jozsa fidelity $\F(\ro,\sig)$ can be a
formidable task. This is in sharp contrast with $\Fn(\ro,\sig)$, which involves only products and traces of density matrices.
Even at the numerical level --- due to the complication involved in evaluating the square root of a Hermitian matrix --- the
computation of $\F(\ro,\sig)$ can be rather resource consuming. For a quantitative understanding of the computational
efficiency, we have performed a numerical comparison of the time required to calculate the fidelity measures $\F$ and $\Fn$,
the nonlogarithmic variety of the quantum Chernoff bound $Q$, and the trace distance $\D$. We have implemented the computations
in both Matlab and C; we present the Matlab codes for reasons of accessibility and succinctness, while the C codes provide
accurate timings without the overhead of the Matlab interpreter.

The time required to evaluate each function was estimated by averaging the times for $100$ pairs of randomly generated
$d$-dimensional density matrices \cite{fn:Random}. Results are shown in Fig. 2 as a function of $d$. The Matlab codes are presented
in Appendix \ref{app:codes}; we attempted to make these codes as efficient as possible within the constraints of the Matlab
environment. Corresponding C codes were implemented as Matlab MEX-files for convenience and can be found online \cite{C:url}.
Our C implementation directly calls the LAPACK and BLAS libraries included in the Matlab distribution for eigenvalue decompositions
and matrix operations. The minimization required in the computation of $Q$ was performed using the Brent minimizer from the GNU
Scientific Library \cite{06Galassi}.

\begin{figure}[!t]
\centering
\includegraphics[width=8cm]{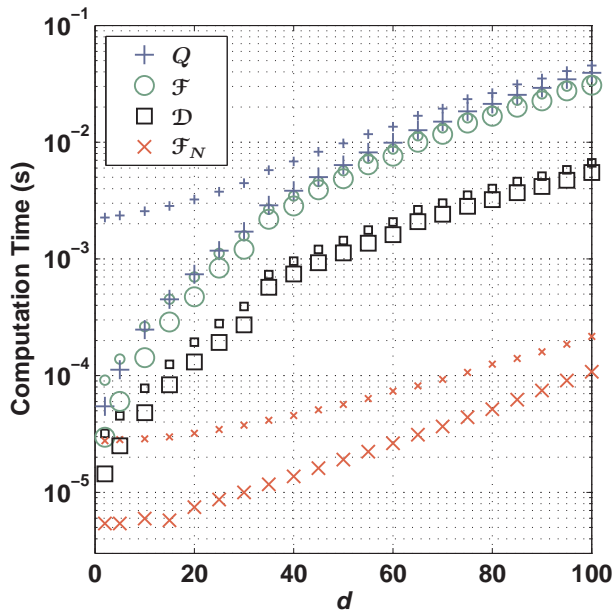}
\caption{(Color online) A semilog plot of the average computation time in Matlab and C for the fidelity measures $\F$ ({$\bigcirc$}),
$\Fn$ ($\times$), the nonlogarithmic variety of the quantum Chernoff bound $Q$ ($+$), and the trace distance $\D$ ($\square$) as a
function of the dimension $d$ of the state space. The smaller and larger markers correspond to timings from Matlab and C, respectively.
Computations were performed on a $2.6$-GHz Intel Pentium 4 CPU.}
\end{figure}

The results shown in Fig. 2 are consistent with the expected algorithmic complexity: Both $\F$ and $Q$ require two Hermitian
diagonalizations, taking an expected $O(d^3)$ operations each \cite{00Parlett38}. Computing $Q$ is the slowest since it requires
both sets of eigenvectors, while $\F$ requires only eigenvalues from one of the diagonalizations. Next fastest is the
computation of $\D$ which requires only eigenvalues from a single diagonalization. All of $\F$, $Q$, and $\D$ appear to require
an asymptotic $O(d^3)$ operation because of the necessity of diagonalization or some other method for computing functions of
the input matrices. On the other hand, our proposed fidelity measure $\Fn$ requires only three Hilbert-Schmidt inner products,
with asymptotic performance $O(d^2)$. Figure 2 clearly shows that the practical numerical evaluation of $\Fn$ is dramatically
faster than the evaluation of $\F$, $\D$, or $Q$. This raises the prospect of using $\Fn$ as a numerically efficient estimate
of distance measures such as $\F$ \cite{08Miszczak} and $\D$ --- particularly for small $d$ where the bounds proven in Sec.
\ref{sec:fntrdistbnd} are tighter. As the dimension increases, the computational advantage of using $\Fn$ becomes even
greater, but the quality of the estimate drops.

\section{Concluding Remarks}\lb{sec:conclusion}

In this paper, we have proposed the quantity $\Fn$ as an alternative fidelity measure (namely, a measure of the degree of
similarity) between an arbitrary pair of mixed quantum states. This measure, the prevailing Uhlmann-Jozsa
fidelity $\F$, and the nonlogarithmic variety of the quantum Chernoff bound $Q$ \cite{07Audenaert160501} are, to the best of
our knowledge, the only known fidelity measures between density matrices that comply with Jozsa's axioms \cite{94Jozsa2315}.
That is, $\F$, $Q$, and $\Fn$ are the only known measures that generalize to pairs of mixed states the concept of fidelity
introduced by Schumacher between a pure and a mixed state \cite{95Schumacher2738}.

The simplicity of $\Fn$ is in sharp contrast with $\F$ and $Q$ since it involves only products of density matrices. Numerically, this
leads to significant reduction in computation time for $\Fn(\ro,\sig)$ over $\F(\ro,\sig)$, especially for higher-dimensional systems.

Besides being easier to compute, $\Fn$ has also been shown to preserve (and even enhance) a number of the useful properties of
$\F$ and $Q$. For example, we have shown that $\Fn$ is a jointly concave measure, that it can be used to place upper and lower
bounds on the value of the trace distance, and that it gives rise to a new metric for the space of density matrices. A
remarkable consequence of the joint concavity of $\Fn$ is that $\F$ is also jointly concave when restricted to a pair of qubit
states --- an interesting problem which remained unsolved thus far \cite{Mike:personal,Uhlmann:personal}.

Our measure, nevertheless, is not without its drawbacks. To begin with, $\Fn$ ---  unlike measures such as $\F$ or $Q$ ---
does not behave monotonically under CPTP maps. In addition, it does not necessarily
vanish when applied to any pair of {\em mixed} states, which are otherwise recognized to be completely different according to
$\F$, $Q$, or their trace distance $\D$. In fact, the explicit dependence on the linear entropies of $\ro$ and $\sig$ gives
rise to the following undesirable feature: the value of $\Fn$ between two completely mixed states residing in disjoint subspaces
can get arbitrarily close to unity as the dimension of the state space tends to infinity.

The undesirable features of $\Fn$ provide a clue as to when $\Fn$ may not be the preferred measure of similarity between two
quantum states: We know that $\Fn$ does not measure the similarity between two high-dimensional, highly mixed states (\ie,
states having non-negligible linear entropy) in the same way that measures like $\F$, $Q$, or $\D$ would. In these cases, the
interpretation of $\Fn$ as a measure of similarity between quantum states must be carried out with extra caution.

With this in mind, we nevertheless see $\Fn$ as an attractive alternative to $\F$. Even when out of its range of applicability, it
follows from a very recent result of Miszczak {\em et al.} \cite{08Miszczak} that $\Fn$ provides an upper bound on the Uhlmann-Jozsa
fidelity $\F$. Moreover, it seems promising that $\Fn$ between any two quantum states may be measured directly in the laboratory,
without resorting to any state tomography protocol \cite{08Miszczak}.

Let us now briefly mention some possibilities for future research that stem from the present work. To begin with, it would be
interesting to search for a quantitative relationship between $\Fn$ and $Q$ analogous to that between $\Fn$ and $\D$
established in this paper or that between $\Fn$ and $\F$ given in Ref.~\cite{08Miszczak}. An estimate of $Q$ based on some
function of $\Fn$ would be useful given that a closed form for $Q$ is not currently known and that $\Fn$ can be computed
relatively easily. In addition, assuming $\Fn$ as an alternative to $\F$, it seems reasonable to reexamine some of the problems
where $\F$ has proven useful, but with $\Fn$ playing its role. In particular, it would be interesting to investigate whether
the simplicity associated with $\Fn$ will offer some advantages over $\F$.

As a first example, we recall from Refs. \cite{97Vedral2275,98Vedral1619} that a standard measure for the amount of
entanglement of a state $\ro$ is given by the shortest {\em distance} from $\ro$ to the set of separable density matrices.
Given the relative simplicity of $\Fn$ with respect to $\F$, it is not inconceivable that a distance measure based on $\Fn$
(such as $C[\Fn]$) may lead to a more efficient determination of this quantity if compared, for example, to $C[\F]$ or the
Bures distance \cite{98Vedral1619}. Of course, since $C[\Fn]$ does not satisfy all the sufficient conditions required to give
rise to a good entanglement measure \cite{97Vedral2275}, any serious attempts in this direction should be preceded by further
investigation of the impact of the nonmonotonicity of $\Fn$ under CPTP maps. For instance, the nonmonotonicity of $\Fn$ might
also imply that the shortest distance from any given state $\ro$ to the set of separable states --- as measured by $C[\Fn]$ ---
does not satisfy the necessary conditions stipulated in Ref.~\cite{00Vidal}, but this is not clear to us at this stage.

As another example, $\Fn$ can be used as a figure of merit in designing optimized quantum control and/or quantum error correction
strategies: One is typically interested in determining a quantum operation $\mathcal{C}$ that minimizes the averaged {\em distance}
between the elements of a set of noisy quantum states $\ro_{i}$ and a predefined set of target quantum states $\sig_{i}$. In this
context, it would be interesting to investigate if distance measures based on $\Fn$ would lead to any advantage in terms of computation
time. Clearly, this has potential applications to the implementation of real-time quantum technologies.

Yet another possible direction of research consists of employing $\Fn$ as a distance measure between quantum operations --- as
opposed to quantum states --- via the isomorphism between quantum states and CPTP maps \cite{99Horodecki1888,99Fujiwara3290}.
In this regard, it is worth investigating whether distance measures based on $\Fn$ would satisfy the six criteria proposed in
Ref. \cite{05Gilchrist062310}. Remarkably, from the results of the present work and Ref. \cite{08Miszczak}, a few strengths of
$\Fn$-based measures can already be anticipated. Of special significance are the fulfillment of the criteria ``easy to
calculate'' and ``easy to measure.'' Along these lines, some operational meaning for $\Fn$ would also be highly desirable.
Although we do not presently have a compelling physical interpretation of $\Fn$, it is not inconceivable that one can be found
in an analogous way to $\F$ \cite{01Spekkens012310}.

\section*{Acknowledgments}

The authors thank Karol \.{Z}yczkowski, Armin Uhlmann and an anonymous referee for useful comments on an earlier version of this
manuscript. P.E.M.F.M. and Y.C.L. acknowledge Sukhwider Singh, Andrew C. Doherty, Alexei Gilchrist, Marco Barbieri, Stephen Bartlett, and Mark
de Burgh for helpful discussions. This work is supported by the Brazilian agencies Coordena\c{c}\~{a}o de Aperfei\c{c}oamento de Pessoal
de N\'{\i}vel Superior (CAPES), Funda\c{c}\~{a}o de Amparo \`{a} Pesquisa do Estado de S\~{a}o Paulo (FAPESP), Project No. 05/04105-5,
the Brazilian Millennium Institute for Quantum Information, Conselho Nacional de Desenvolvimento Cient\'{\i}fico e Tecnol\'{o}gico
(CNPq), and the Australian Research Council.

\appendix
\section{Metrics}\lb{app:metrics}

From a mathematically rigorous viewpoint, a distance measure $\mathfrak{D}$ on a set $S$ is a function $\mathfrak{D}: S \times
S \to \mathbb{R}$ such that for every $ a,b,c \in S$ the following properties hold.
\begin{enumerate}
\item[(M1)] $\mathfrak{D}(a,b) \geq 0$ (non-negativity),
\item[(M2)] $\mathfrak{D}(a,b) = 0$ iff $a = b$ (identity of indiscernible),
\item[(M3)] $\mathfrak{D}(a,b) = \mathfrak{D}(b,a)$ (symmetry),
\item[(M4)] $\mathfrak{D}(a,c) \leq \mathfrak{D}(a,b) + \mathfrak{D}(b,c)$ (triangle inequality).
\end{enumerate}
Any such function is called a {\em metric}.

\section{Proofs}
\subsection{Proof of Proposition \ref{prop:concFx}}\lb{app:concFx}

In this appendix the joint concavity of $\Fn$ is established via the proof of Proposition \ref{prop:concFx}.
\begin{proof}
Differentiating Eq. \eqref{eq:Fx} twice with respect to $x$, we obtain
\brr
\lb{eq:Fder2}
\frac{d^{2} F(x)}{dx^{2}} &=& 2 \vec{u} \cdot \vec{v} + \frac{d^{2} f(x)}{dx^{2}} g(x) + f(x) \frac{d^{2}g(x)}{dx^{2}} \nn \\
& & + \, 2 \frac{d f(x)}{dx} \frac{d g(x)}{dx}\,,
\err
where, for convenience, we define the functions $f(x) \defeq \sqrt{1 - \| \vec{r} + x \vec{u} \|^{2}}$ and $g(x) \defeq \sqrt{1 - \|
\vec{s} + x \vec{v} \|^{2}}$. After some computation we find that
\be
\frac{d^{2} F(x)}{dx^{2}} = \mathfrak{F}_{1}(x) + \mathfrak{F}_{2}(x) ,
\ee
where
\brr
\!\!\!\!\!\!\!\!\!\!\! & & \mathfrak{F}_{1}(x) \defeq 2 \vec{u} \cdot \vec{v} - \frac{g(x) u^{2}}{f(x)} - \frac{f(x) v^{2}}{g(x)} , \\
\!\!\!\!\!\!\!\!\!\!\! & & \mathfrak{F}_{2}(x) \defeq 2 \, \frac{\vec{u} \cdot (\vec{r} + x \vec{u}) \vec{v} \cdot (\vec{s} + x
\vec{v})}{f(x) g(x)} \nn \\
\!\!\!\!\!\!\!\! & & - \frac{g(x) [ \vec{u} \cdot (\vec{r} + x \vec{u}) ]^{2}}{[f(x)]^{3}} - \frac{f(x) [ \vec{v} \cdot
(\vec{s} + x \vec{v}) ]^{2}}{[g(x)]^{3}} . \err
The negative semidefiniteness of $d^{2} F(x) / dx^{2}$ in the range $x \in [0,1]$ can be observed if $\mathfrak{F}_{1}(x)$ and
$\mathfrak{F}_{2}(x)$ are written in the following alternative form:
\brr
\mathfrak{F}_{1}(x) &=& - \left\| \sqrt{\frac{g(x)}{f(x)}} \, \vec{u} - \sqrt{\frac{f(x)}{g(x)}} \, \vec{v} \right\|^{2} , \nn \\
\mathfrak{F}_{2}(x) &=& - \frac{1}{f(x) g(x)} \nn \\
& & \times \lbk \frac{g(x)}{f(x)} \vec{u} \cdot (\vec{r} + x \vec{u}) - \frac{f(x)}{g(x)} \vec{v} \cdot (\vec{s} + x \vec{v})
\rbk^{2} . \nn
\err
\end{proof}

\subsection{Proof of supermultiplicativity of $\Fn$}
\lb{app:super-multplicativity}

To prove that $\Fn$ is supermultiplicative, we first define $r_{i} \defeq \tr (\ro_{i}^{2})$ and $s_{i} \defeq \tr
(\sig_{i}^{2})$, such that $0 < r_{i}, s_{i}\leq 1$ (note that here we use $r_{i}$ instead of $r_{i}^{2}$ as the norm square
of $\vec{r}_{i}$, likewise for $s_{i}$). Straightforward algebra gives
\begin{gather*}
\Fn(\ro_{1} \otimes \ro_{2},\sig_{1} \otimes \sig_{2}) - \Fn(\ro_{1},\sig_{1}) \Fn(\ro_{2},\sig_{2}) = \\
\sqrt{(1 - r_{1} r_{2})(1 - s_{1} s_{2})} \\
- \sqrt{(1 - r_{1})(1 - s_{1})(1 - r_{2})(1 - s_{2})} \\
- \tr (\ro_{1} \sig_{1}) \sqrt{(1 - r_{2})(1 - s_{2})} \\
- \tr (\ro_{2} \sig_{2}) \sqrt{(1 - r_{1})(1 - s_{1})} .
\end{gather*}
A direct application of Cauchy-Schwarz's inequality $\tr (\ro_{i} \sig_{i}) \leq \sqrt{r_{i} s_{i}}$ gives
\begin{gather*}
\Fn(\ro_{1} \otimes \ro_{2},\sig_{1} \otimes \sig_{2})- \Fn(\ro_{1},\sig_{1}) \Fn(\ro_{2},\sig_{2}) \geq \\
\sqrt{(1 - r_{1} r_{2})(1 - s_{1} s_{2})} \\
- \sqrt{(1 - r_{1})(1 - s_{1})(1 - r_{2})(1 - s_{2})} \\
- \sqrt{r_{1} s_{1} (1 - r_{2})(1 - s_{2})} - \sqrt{r_{2} s_{2} (1 - r_{1})(1 - s_{1})} .
\end{gather*}

The supermultiplicative property is obtained by showing the positive semidefiniteness of the right-hand side of the above expression. This is
the content of the following proposition.
\begin{proposition}
For $0 \leq a,b,c,d \leq 1$, we have
\begin{multline}
\lb{eq:supermult}
\sqrt{(1-ab)(1-cd)} \geq \sqrt{(1-a)(1-b)(1-c)(1-d)} \\
\!\!\! + \sqrt{ac(1-b)(1-d)} + \sqrt{bd(1-a)(1-c)} .
\end{multline}
\end{proposition}
\begin{proof}
First note that if any of the variables equals $1$, then the validity of the inequality is immediate. For example, let $d=1$
so that inequality \eqref{eq:supermult} reduces to
\be
\sqrt{(1-a b)(1-c)} \geq \sqrt{b(1-a)(1-c)} .
\ee
This is trivially satisfied for all $0 \leq a,b,c\leq 1$. In what follows, we restrict ourselves to $0 \leq a,b,c,d < 1$ and show that
inequality \eqref{eq:supermult} is equivalent to the standard inequality of arithmetic and geometric means (hereafter referred
as the AM-GM inequality). This inequality is just an expression of the fact that the geometric mean of a list of non-negative
real numbers is never larger than the corresponding arithmetic mean.

Apply the substitution $a^{\prime} = 1 - a$ (similarly for $b^{\prime}$, $c^{\prime}$, and $d^{\prime}$; note that $0 <
a^{\prime}, b^{\prime},c^{\prime},d^{\prime} \leq 1$) to inequality \eqref{eq:supermult} and divide the result by
$\sqrt{a^{\prime} b^{\prime} c^{\prime} d^{\prime}}$ to get the equivalent inequality
\be
\lb{eq:ineqsupermult}
\! \sqrt{(1+A+B)(1+C+D)} \geq 1 + \sqrt{AC} + \sqrt{BD} ,
\ee
where we have defined $A = 1/a^{\prime} - 1$ (similarly for $B$, $C$, and $D$; note that $0 \leq A,B,C,D < \infty$). Squaring
the inequality above we find
\be
\! \frac{A\!+\!C}{2} + \frac{B\!+\!D}{2} + \frac{AD\!+\!BC}{2} \geq \sqrt{AC} \!+\! \sqrt{BD} \!+\! \sqrt{ABCD}
\ee
which is clearly a sum of three AM-GM inequalities.
\end{proof}

\subsection{Proof of the metric property of $B[\F]$ and $C[\F]$}\lb{app:metric_C}

In the following, we give an alternative demonstration of the metric properties of $B[\F]$ and $C[\F]$ (see Refs. \cite{69Bures199,05Gilchrist062310}
for the standard proofs). Our proof consists of a simple application of Theorem \ref{thm:schoenberg} due to Schoenberg.
\begin{proposition}
\lb{prop:CFmetric} The functionals $B[\F(\ro,\sig)]$ and $C[\F(\ro,\sig)]$, defined in Eq.~\eqref{eq:BF} and Eq.~\eqref{eq:CF},
are metrics for the space of density matrices.
\end{proposition}
\begin{proof}
Let $K[\F(\ro,\sig)]$ represent either $B[\F(\ro,\sig)]$ or $C[\F(\ro,\sig)]$ for brevity. As with $\F$, it is easy to check that
$K^{2}[\F(\ro,\sig)]$ is symmetric in its two arguments and that $K^{2}[\F(\ro,\sig)] \geq 0$ with saturation iff $\ro = \sig$. So,
according to Theorem \ref{thm:schoenberg}, $K[\F(\ro,\sig)]$ is a metric if for any set of density matrices $\{ \ro_{i} \}_{i=1}^{n}$
($n \geq 2$) and real numbers $\{ c_{i} \}_{i=1}^{n}$ such that $\sum_{i=1}^{n} c_{i} = 0$, it is true that
\be
\lb{eq:toprove}
\sum_{i,j = 1}^{n} K^{2}[\F(\ro_{i},\ro_{j})] \, c_{i} c_{j} \leq 0 .
\ee
To prove this, we derive an upper bound for $K^{2}[\F(\ro_{i},\ro_{j})]$, which can be easily seen to satisfy the condition above. First,
note that
\begin{multline}
\lb{eq:seqofineqs}
\F(\ro_{i},\ro_{j}) = \lbk \tr \lpar | \sqrt{\ro_{i}} \sqrt{\ro_{j}}| \rpar \rbk^{2} \geq \left| \tr \lpar \sqrt{\ro_{i}} \sqrt{\ro_{j}}
\rpar \right|^2 =  \\
\lbk \tr \lpar \sqrt{\ro_{i}} \sqrt{\ro_{j}} \rpar \rbk^{2} = \tr \lpar \sqrt{\ro_{i}} \sqrt{\ro_{j}} \otimes \sqrt{\ro_{i}}
\sqrt{\ro_{j}} \rpar = \\
\tr \lbk \lpar \sqrt{\ro_{i}} \otimes \sqrt{\ro_{i}} \rpar \lpar \sqrt{\ro_{j}} \otimes \sqrt{\ro_{j}} \rpar \rbk \equiv
\mathcal{A}(\ro_{i},\ro_{j}) ,
\end{multline}
where the first equality follows from the definition $| {\bf A} | \defeq \sqrt{{\bf A}^{\dagger} {\bf A}}$ for every matrix ${\bf A}$
and the inequality from the fact that $\tr (| {\bf A} |) = \max_{{\bf U}} | \tr ({\bf UA}) |$ (the maximization runs over unitary
matrices ${\bf U}$ \cite{94Jozsa2315,60Schatten}). Then, it follows that
\begin{align}
B^{2}[\F] & = 2 (1 - \sqrt{\F}) \leq 2 (1 - \F) \leq 2 (1 - \mathcal{A}) , \\
C^{2}[\F] & = 1 - \F \leq 1 - \mathcal{A} \leq 2 (1 - \mathcal{A}) ,
\end{align}
or, in our more compact notation, $K^{2}[\F] \leq 2 (1 - \mathcal{A})$.

Now, replacing $K^{2}[\F(\ro_{i},\ro_{j})]$ with the above upper bound in the left-hand side of Eq. \eqref{eq:toprove}, it is easy to obtain the
desired inequality:
\begin{multline}
\sum_{i,j=1}^{n} \lbr 2 - 2 \tr \lbk \lpar \sqrt{\ro_{i}} \otimes \sqrt{\ro_{i}} \rpar \lpar \sqrt{\ro_{j}} \otimes \sqrt{\ro_{j}}
\rpar \rbk \rbr c_{i} c_{j} = \\
\!\!\!\!\! - 2 \tr \lpar \left| \sum_{i=1}^{n} c_{i} \sqrt{\ro_{i}} \otimes \sqrt{\ro_{i}} \right|^2 \rpar \leq 0 ,
\end{multline}
where the equality is obtained by using the fact that $\sum_{i=1}^{n} c_{i} = 0$, the linearity of the trace operation, and the
hermiticity of $c_{i} \sqrt{\ro_{i}} \otimes \sqrt{\ro_{i}}$.
\end{proof}

Finally, let us just mention that besides establishing the metric properties of $B[\F]$ and $C[\F]$, the present proof also establishes
$\sqrt{2 - 2 \lbk \tr \lpar \sqrt{\ro} \sqrt{\sig} \rpar \rbk^{2}}$ as a metric for the space of density matrices. In fact, by a
similar application of Schoenberg's theorem, the quantity $H(\ro,\sig) \defeq \sqrt{2 - 2 \tr \lpar \sqrt{\ro} \sqrt{\sig} \rpar}$ can
also be shown to be a metric.

\section{Matlab Codes}\lb{app:codes}

In this appendix, we present the Matlab codes that we have used to compute the various functions involved in the numerical experiment
presented in Sec. \ref{sec:computability}.

For \verb|rho| and \verb|sigma| density matrices,
\begin{itemize}
\item $\Fn$ was computed using
\begin{verbatim}
Fn = real( rho(:)'*sigma(:) ...
     + sqrt((1 - rho(:)'*rho(:))* ...
     (1 - sigma(:)'*sigma(:))) );
\end{verbatim}

\item $\F$ was computed using
\begin{verbatim}
[V, D] = eig(rho);
sqrtRho = V*diag(sqrt(diag(D)))*V';
F = sum( sqrt(eig(Hermitize( ...
         sqrtRho*sigma*sqrtRho))) )^2;
\end{verbatim}
Here \verb|sqrtRho*sigma*sqrtRho| is not quite Hermitian due to small numerical errors. We therefore employ the function
\verb|Hermitize(M)=(M+M')/2| to turn the almost-Hermitian matrix into a Hermitian one --- this causes Matlab to select a
more efficient algorithm for the diagonalization.

\item $\D$ was computed using
\begin{verbatim}
    D=0.5*sum(abs( eig(rho-sigma) ));
\end{verbatim}

\item $Q$ was computed using
\begin{verbatim}
[Vr,Drho]=eig(rho); Dr=diag(Drho);
[Vs,Dsigma]=eig(sigma); Ds=diag(Dsigma);
A = abs(Vr'*Vs).^2;
[x,Q]=fminbnd(@(s) ...
      (Dr.'.^s)*A*(Ds.^(1-s)), 0, 1);
\end{verbatim}
The algorithm used here follows from the formula for $\tr(\ro^{s} \sig^{1-s})$ given in the section entitled {\em convexity in s} of
Ref. \cite{07Audenaert160501}.
\end{itemize}


\end{document}